\documentclass[twocolumn,prl,nofootinbib,unsortedaddress,superscriptaddress,showpacs]{revtex4}

\usepackage{graphicx}
\usepackage{dcolumn}
\usepackage{amsmath}
\usepackage{url}
\usepackage{bbm}

\newcommand{\id}{\mathbbm{1}}

\begin{document}

\title{Discreteness of area in noncommutative space}

\author{Giovanni AMELINO-CAMELIA}
\email{giovanni.amelino-camelia@roma1.infn.it}
\affiliation{Dipartimento di Fisica \\
 Universit\`a di Roma ``La Sapienza"\\
and Sez.~Roma1 INFN\\
P.le A. Moro 2, 00185 Roma , Italy}

\author{Giulia GUBITOSI}
\affiliation{Dipartimento di Fisica \\
 Universit\`a di Roma ``La Sapienza"\\
and Sez.~Roma1 INFN\\
P.le A. Moro 2, 00185 Roma , Italy}

\author{Flavio MERCATI}
\affiliation{Dipartimento di Fisica \\
 Universit\`a di Roma ``La Sapienza"\\
and Sez.~Roma1 INFN\\
P.le A. Moro 2, 00185 Roma , Italy}

\begin{abstract}
We introduce an area operator for the Moyal noncommutative plane.
We find that the spectrum is discrete, but, contrary to the expectation
formulated by other authors, not characterized
by a ``minimum-area principle". We show that an intuitive analysis
of the uncertainty relations obtained from Moyal-plane noncommutativity
is fully consistent with our results for the spectrum, and
we argue that our area operator  should be generalizable to several other
noncommutative spaces. We also observe that the properties of distances and areas
in the Moyal plane expose some weaknesses in the line of reasoning adopted in some of the
heuristic analyses of the measurability of geometric spacetime observables in
the quantum-gravity realm.
\end{abstract}

\pacs{}

\maketitle

Motivation for the study of spacetime noncommutativity comes primarily
from its possible use in investigations of spacetime fuzziness,
which in particular  is expected~\cite{mead,padma,ngmpla,gacmpla,ahluGUP,garay}
to be relevant for the description
of the quantum-gravity realm.
However, most work focuses on establishing how
a noncommutativity of spacetime would affect
S-matrix/field-theory observables
(such as  the probabilities of occurrence of certain particle-physics processes)
and still very little has been actually established concerning
the fuzziness and other properties of geometric observables.
While in Loop Quantum Gravity~\cite{lqg1,lqg2,lqg3,lqg4},
another much-studied formalism with a quantum geometry,
there is a deservedly renowned detailed analysis~\cite{areaLQG}
of the properties of areas and volumes,
very little is known about areas and volumes
even for the simplest noncommutative spacetimes, the ``canonical spacetimes"~\cite{moyalrefs3}
characterized by noncommutativity of the type ($\{ \mu,\nu \} \in \{ 0,1,2,3 \}$)
\begin{equation}
\left[ {\hat X}_\mu, {\hat X}_\nu \right]=i \theta_{\mu\nu}~
\label{canonical}
\end{equation}
(with a noncommutativity matrix $\theta_{\mu \nu}$ that
 commutes\footnote{While over the last few years canonical spacetimes are indeed mostly
 studied assuming that $\theta_{\mu \nu}$
 commutes with the coordinates, the research programme that first led to the proposal
 of canonical noncommutativity~\cite{dopli1994} is contemplating the possibility~\cite{doplithetanontrivi}
 of a $\theta_{\mu \nu}$ that does not commute with the coordinates.}
  with the coordinates $X_\mu$).

   To our knowledge, the most explicit (and yet only tentative) investigation
  of any of these issues is the one of Ref.~\cite{romero}, which reported an attempt
  of characterization of the areas of discs in the (Groenewold-)Moyal
  plane~\cite{moyalrefs1,moyalrefs2},
which is the case of canonical noncommutativity for only two spatial dimensions:
\begin{equation}
\left[ {\hat X}_1, {\hat X}_2\right]=i \theta~.
\label{moyal}
\end{equation}

We shall here take as starting point
the strategy of analysis advocated in Ref.~\cite{romero},
but provide a more satisfactory characterization of areas in the Moyal plane,
which ends up showing that, while some aspects of the methodology
proposed  in Ref.~\cite{romero} are indeed very valuable,
the specific indications concerning the properties of areas
that emerged in the study reported  in Ref.~\cite{romero}
were misleading.

It is convenient for us to start
with a brief description of the
analysis reported in Ref.~\cite{romero}. And we must immediately
underline a first potentially worrisome aspect of that
analysis concerning the fact that discs in the noncommutative plane
were specified~\cite{romero} through the coordinates
of a single point.
The area ${\hat {\cal A}}$ of the disc was
described by generalizing a familiar commutative-spacetime formula
\begin{equation}
 \hat {\cal A}=\pi\left[ {\hat X}_1^2 + {\hat X}_2^2\right]. \label{eq:arearomero}
\end{equation}
in terms of the noncommutative coordinates
of that single point.
So the discs considered in Ref.~\cite{romero} are centered around
a ``classical origin" (but in the Moyal plane no point, not even one chosen as
the origin, can truly be treated classically) and their boundaries are identified
by a single point with coordinates ${\hat X}_1, {\hat X}_2$.

Ref.~\cite{romero} studies the spectrum of $\hat {\cal A}$, on the basis of
the fact that  ${\hat  X}_1, {\hat X}_2$ are governed by (\ref{moyal}), and finds that
\begin{equation}
 {\cal A}_n=2\pi\theta(n+\frac{1}{2}),
\end{equation}
which would amount to a discrete spectrum with ``quanta of area" of $2\pi\theta$
and with a minimum eigenvalue $\pi\theta$.

Our motivation in setting up a more refined investigation of areas in the Moyal plane
came from the observation that it seemed to us unreasonable that
the Moyal-plane noncommutativity would give rise to a ``minimum area principle"
(independently of whether the minimum value was $\pi\theta$ or some other value).
From (\ref{moyal}) one infers
uncertainty relations of the type
\begin{equation}
\delta x_1 \delta x_2 \geq |\theta|/2 ~,\label{moyalUncertain}
\end{equation}
which do not prevent points from having ``sharp" value of one of the coordinates,
although this comes at the cost of uncontrolled uncertainties in the other coordinate.
Therefore it should be possible to have situations in which
sharply (of course, in an appropriate ``state"
of the Moyal quantum geometry~\cite{agmStringspace}) all of
the  boundary points that identify a surface
have the same value of, say, $x_1$, which is the case in which the surface collapses sharply to
having zero area.

This bit of intuition appeared to render more significant what could have been viewed
as mere
technical limitations of the characterization of area offered in Ref.~\cite{romero}:
(I) the formula for ${\hat {\cal A}}$ makes reference (implicit but substantial) to a classical
center of the disc, which, as stressed above, is a concept that is clearly foreign
to the Moyal-plane  quantum geometry, and (II) the formula for ${\hat {\cal A}}$  codifies
the information on the boundary of the surface through the coordinates of
a single point, which is the highest conceivable level of optimism (specifying
a boundary with less than one point
would require more than optimism).

On the basis of these considerations (both intuitively originating
from $\delta x_1 \delta x_2 \geq |\theta|/2 $ and technically originating from the
structural inadequacies, as a candidate area operator, of the operator ${\hat {\cal A}}$)
we concluded that ${\hat {\cal A}}$ of Ref.~\cite{romero} cannot be a meaningful
tool of investigation of the spectrum of areas.
And in devising an alternative strategy we found guidance also in the observation that even in
elementary commutative geometry one could reasonably argue~\cite{klein}
that areas are most transparently characterized for polygons  with $N$ vertices, with the case
of surfaces with smooth boundaries (such as discs) discussed as a careful $N \rightarrow \infty $ limit
in which the differential calculus plays a central role.
In noncommutative spaces one does have a notion of differential calculus~\cite{majidbook},
but not immune from certain counter-intuitive peculiarities (see, {\it e.g.}, Ref.~\cite{sitarz}),
which in turn may affect nontrivially the description of surfaces with smooth boundaries
as a limit $N \rightarrow \infty $ of polygons.
Moreover, the Loop-Quantum-Gravity results on the spectrum of volume in a 3-dimensional spatial
quantum geometry guided us to expect that triangles would have to be the most
natural framework for area investigation in the Moyal plane: in the description of volumes
in the 3-dimensional quantum geometry of Loop Quantum Gravity tetrahedra play a pivotal
 role~\cite{lqg1,areaLQG},
which we may well expect to be similar to the role played by triangles in the description
of areas in two-dimensional
quantum geometries, such as the Moyal plane.

We therefore set out to establish whether or not the spectrum of areas is discrete in the
Moyal plane, and whether or not this spectrum is characterized by a ``minimum-area principle"
(a minimum, nonzero, allowed sharp value of area), by focusing primarily on triangles.
The starting point for our proposal of a triangle area operator for the Moyal plane
is an elementary formula
in commutative spacetime (whose points have commutative coordinates $ \xi_1,\xi_2$)
which describes~\cite{klein}
the area of a triangle with
vertices ${\vec \xi}^{\, (1)} \equiv \{\xi_1^{(1)},\xi_2^{(1)}\}$,
 ${\vec \xi}^{\, (2)} \equiv \{\xi_1^{(2)},\xi_2^{(2)}\}$
and  ${\vec \xi}^{\, (3)} \equiv \{\xi_1^{(3)},\xi_2^{(3)}\}$
in the following way
\begin{equation}
A^{(triangle)} = |A({\vec \xi}^{\,\, (1)},{\vec \xi}^{\,\, (2)},{\vec \xi}^{\,\, (3)})|
\end{equation}
in terms of
\begin{equation}
 A({\vec \xi}^{\,\, (1)},{\vec \xi}^{\,\, (2)},{\vec \xi}^{\,\, (3)}) = \frac{1}{2}\,det \left[
\begin{array}{lll}
 \xi_1^{(1)}& \xi_2^{(1)} & 1 \\
\xi_1^{(2)} & \xi_2^{(2)} & 1 \\
\xi_1^{(3)} & \xi_2^{(3)} & 1
                    \end{array}\right]
                    ~.
                    \label{areaclassic}
\end{equation}
The formula
 (\ref{areaclassic}) is exactly of the type we should deem desirable for the purposes
 of proposing a quantum version. It depends exclusively on the coordinates of the vertices,
 without any reference to ``special" external points (such as an origin, which might then
 generate embarrassment in the quantum version, in the sense discussed above) .
And the fact that this formulation as the
absolute value of  $A({\vec \xi}^{\, (1)},{\vec \xi}^{\, (2)},{\vec \xi}^{\, (3)})$
expresses the area as a function of the coordinates of the vertices in a way
that is invariant under permutations of the vertices
allows us to avoid a potential difficulty: formulas that depend on identifying the sequence
of vertices would be unpleasant in a fuzzy geometry, since in principle the fuzziness should
prevent us from being able to establish such sequences in general.

In light of these observations we promote $A({\vec \xi}^{\, (1)},{\vec \xi}^{\, (2)},{\vec \xi}^{\, (3)})$
to the status of a quantum-geometry
observable, obtained from the quantum coordinates of three points in the Moyal plane,
by posing
\begin{equation}
{\hat A}({\vec {\hat X}}^{\, (1)},{\vec{\hat X}}^{\, (2)},{\vec {\hat X}}^{\, (3)}) \equiv \frac{1}{2}\,det \left[
\begin{array}{lll}
 \hat X_1^{(1)}& \hat X_2^{(1)} & 1 \\
\hat X_1^{(2)} & \hat X_2^{(2)} & 1 \\
\hat X_1^{(3)} & \hat X_2^{(3)} & 1
                    \end{array}\right]
                    ~,
\end{equation}
and we shall describe the area ${\hat A}^{(triangle)}$ of a triangle in the Moyal plane as the absolute value of
the expectation
of ${\hat A}({\vec {\hat X}}^{\, (1)},{\vec{\hat X}}^{\, (2)},{\vec {\hat X}}^{\, (3)})$ in a quantum state of
the Moyal-plane geometry.

We introduced a dedicated notation $X_j^{(m)}$ for the j-th coordinate of the m-th point
also as a reminder of the fact that in order to consider at once three points
in a noncommutative space a couple of steps of formalization must be taken.
The relevant issues may be discussed rigorously at the abstract algebraic level
(see, {\it e.g.}, Ref.~\cite{majidbook}),
but as physicists we find particularly comfortable to rely on representations. ${\hat X}_1$
and ${\hat X}_2$ in (\ref{moyal}) can be described as operators on a Hilbert space~\cite{majidbook,gacmajid}
with structure that exactly reproduces the Hilbert space of a particle in nonrelativistic
quantum mechanics. Denoting the ``state of the point" by $|\psi>$, so that
 the  ``wave functions" of the first coordinate of the point is $\psi(x_1)=<x_1|\psi>$, in light
 of (\ref{moyal}) we can prescribe that ${\hat X}_1$
and ${\hat X}_2$ act as follows
\begin{equation}
{\hat X}_1 \triangleright \psi(x_1)
= x_1 \, \psi(x_1)
 ~ ,
\label{actionphi}
\end{equation}
\begin{equation}
{\hat X}_2 \triangleright  \psi(x_1)
= - i \theta \frac{\partial}{\partial x_1} \psi(x_1)
  ~ .
\label{actionpi}
\end{equation}
Our observable ${\hat A}({\vec {\hat X}}^{\, (1)},{\vec{\hat X}}^{\, (2)},{\vec {\hat X}}^{\, (3)})$
is a 3-point observable, so it
acts on states $|\psi\! \! \! \! >$ such
that $\left(<x_1^{(1)}|\otimes <x_1^{(2)}|\otimes <x_1^{(3)}| \right) |\psi \! \! \! \!
>=\psi(x^{(1)}_1,x^{(2)}_1,x^{(3)}_1)$.
And the operators ${\hat X}_i^{\, (1)}$, ${\hat X}_i^{\, (2)}$, ${\hat X}_i^{\, (3)}$
are of course intended as follows:
\begin{eqnarray}
 \hat X_i^{(1)}\equiv \hat X_i\otimes \id\otimes \id,\;\;&&
 \hat X_i^{(2)}\equiv \id \otimes \hat X_i \otimes \id ,\;\;\nonumber\cr
 \hat X_i^{(3)}\equiv \id \otimes  \id \otimes \hat X_i  ~,
\end{eqnarray}
where $i=1,2$.

These clarifications, besides giving a precise characterization of the observable $\hat A$,
also set the stage for  a rather straightforward derivation of the spectrum of $\hat A$.
For this purpose it is convenient to introduce the notation $\hat q_i$ for the first coordinate
of the i-th point and $ \hat p_i$ for
the second coordinate
of the i-th point
\begin{equation}
 \hat q_i \equiv \hat X_1^{(i)}~,~~~ \hat p_i \equiv \hat X_2^{(i)} ~.
\end{equation}
This allows  us to rewrite $\hat A$ as follows
\begin{equation}
 \hat A = \frac{1}{2} \Sigma_j \epsilon^{jkl} \hat q_k \hat p_l ~,
\end{equation}
and the correspondence with a Hamiltonian interaction term which is familiar
to physicists becomes evident upon noticing that
\begin{equation}
\left[\hat q_i , \hat p_j\right]=i \theta \delta_{ij}~,
\label{thetaHeisen}
\end{equation}
since
from (\ref{moyal}) it follows
that $ \left[ \hat X_i^{(j)} , \hat X_k^{(l)} \right]=i \theta(1- \delta_{ik})\delta^{jl}$.
Clearly $\hat A$ has all the properties of the familiar operator $\frac{1}{2} \Sigma_j \hat{J}_j  B_j$
describing the interaction between the angular momentum observable
and an external (classical) homogeneous magnetic field $\vec B=\left(1,1,1\right)$,
but with the Planck constant denoted by $\theta$ (see (\ref{thetaHeisen})).
This allows us to deduce that the spectrum of $\hat A$
is  $A_n=\sqrt 3 \theta n/2$, with $n\in \mathbbm Z$, and therefore the spectrum of
the area of the triangle is
\begin{equation}
 A^{(triangle)}_n=\frac{\sqrt 3}{2} \,  |\theta n| ~, \label{spectrA}
\end{equation}
We conclude that in the Moyal plane the allowed sharp values of area of a triangle
are characterized by a regular discretization
with ``quantum of area" $\frac{\sqrt 3}{2}|\theta|$,
and most importantly the Moyal plane is {\underline{not}}
an example of quantum space in which a ``minimum area principle"
holds: the lowest eigenvalue of ${\hat A}^{(triangle)}$ is $A_0=0$.
And in particular, as expected (see the considerations guided by $\delta x_1 \delta x_2 \geq |\theta|/2$
which we offered above), our area of the triangle in the Moyal noncommutative plane
vanishes if the three vertices
have the same $x_1$ coordinates. In fact, it is easy to verify that
\begin{eqnarray}
  {\hat A}({\vec {\hat X}}^{\, (1)},{\vec{\hat X}}^{\, (2)},{\vec {\hat X}}^{\, (3)})
|\psi_0 >=0
\label{jocL1}
\end{eqnarray}
if
\begin{eqnarray}
\psi_0(x^{(1)}_1,x^{(2)}_1,x^{(3)}_1)
= \delta(x_1^{(1)}-x^*)\delta(x_1^{(2)}-x^*)\delta(x_1^{(3)}-x^*) \nonumber\cr
\label{jocL2}
\end{eqnarray}

Concerning our intuition that triangles should play a role in the general description of
areas in the Moyal plane that is just as central as the role of tetrahedra in the Loop-Quantum-Gravity
description of general volumes (perhaps a different role, but equally central),
we can only report some very preliminary observations, which however appear to confirm
our expectations.
A starting point is provided by the (very elementary) observation
that in an ordinary (commutative) plane the areas of all polygons
can be obtained as a sum of areas of triangles.
And a formula giving the area of the polygon
as function of the coordinates of the vertices can be arranged~\cite{klein}
as a sum of functions of the type  $A({\vec \xi}^{\,\, (l)},{\vec \xi}^{\,\, (m)},{\vec \xi}^{\,\, (n)})$,
which we gave in our Eq.~(\ref{areaclassic}) (and then promoted to the status of
quantum operator ${\hat A}({\vec {\hat X}}^{\, (l)},{\vec{\hat X}}^{\, (m)},{\vec {\hat X}}^{\, (n)})$
in the Moyal plane).
What changes is that for polygons with $N\ge 4$ vertices the formulas
giving the area of the polygon as a sum of the areas of triangles always genuinely
depend on the ordering
of the vertices~\cite{klein}.
The independence on the ordering of vertices, which, as we stressed, is available for
triangles,  played a crucial role in our analysis of the spectrum of the area of triangles, and without
it clearly some new tool of analysis must be introduced. But even before having such a new tool
we can already provide evidence that there will be polygons with sharply zero area.
This we see of course from the fact that
when the state of the vertices of the polygon is of the type
\begin{eqnarray}
\psi_0(x^{(1)}_1,x^{(2)}_1,...,x^{(N)}_1) = \delta(x_1^{(1)} \! \! \! - \! x^*)
 \delta(x_1^{(2)} \! \! \! - \! x^*)...\delta(x_1^{(N)} \! \! \! - \! x^*)
\nonumber
\end{eqnarray}
then for each triplet of vertices (labeled $l$, $m$, $n$) of the $N$-vertices polygon one has
\begin{eqnarray}
  {\hat A}({\vec {\hat X}}^{\, (l)},{\vec{\hat X}}^{\, (m)},{\vec {\hat X}}^{\, (n)})
|\psi_0 >=0
~.
\nonumber
\end{eqnarray}
So, even though, as stressed above,
for $N \ge 4$ the relationship between the area and a certain sum of terms of the
type  ${\hat A}({\vec {\hat X}}^{\, (l)},{\vec{\hat X}}^{\, (m)},{\vec {\hat X}}^{\, (n)})$
requires a corresponding choice of ordering of the vertices, it seems we can conclude
that the area is zero on $|\psi_0 >$, since
each term in the sum
vanishes on $|\psi_0 >$, independently of the ordering of the vertices.

Both in our detailed analysis of the area of triangles and in these brief
preliminary remarks on general polygons we
placed particular significance on the existence of zero-area states
because this feature (already rather significant intrinsically within
the exploration of the Moyal plane)
has some broader implications. To see this
we must first briefly summarize the simpler result on the
squared-distance observable which we reported in Ref.~\cite{agmStringspace} (within
an analysis of distance observables in a few different examples of quantum spaces).
Unsurprisingly in Ref.~\cite{agmStringspace} we adopted as squared-distance observable
the operator
\begin{equation}
 {\hat d^2} \equiv \left( \hat X_1^{(1)} -  \hat X_1^{(2)}\right)^2
 +\left(\hat X_2^{(1)}-\hat X_2^{(2)}\right)^2~,
\end{equation}
where $\hat X_i^{(1)}\equiv \hat X_i\otimes \id$, $\hat X_i^{(2)}\equiv \id \otimes \hat X_i$.

Introducing here $\hat q \equiv \hat X_1^{(1)}-\hat X_1^{(2)}$ and $\hat p \equiv \hat X_2^{(1)}-\hat X_2^{(2)}$
one of course can rewrite ${\hat d^2}$ as follows
\begin{equation}
{\hat d^2} = \frac{\hat p^2}{2 { m}}+\frac{1}{2}m \omega^2 \hat q^2,
\end{equation}
for ${ m}=\frac{1}{2}$, $\omega=2$. And observing that
 from (\ref{moyal}) it follows
that $\left[\hat q,\hat p\right]=2i \theta$,
one easily concludes~\cite{agmStringspace}
that the spectrum of $d^2$ is of the harmonic-oscillator type:
\begin{equation}
 d^2_j = 4 |\theta| \left( j +\frac{1}{2} \right)~,
\end{equation}
with $j$ integer and nonnegative.

So also for ${\hat d^2}$ one finds a discrete spectrum, with quanta of squared-distance
of $4|\theta|$, but interestingly here there is a ``minimum-distance principle" at work:
the lowest eigenvalue is  $d^2_0=2 |\theta|$.

We conclude that the Moyal plane is an example of noncommutative space
in which there is a minimum distance but no minimum area.
A posteriori one can see that this should have been expected, considering
once again the uncertainty relations implied
by Moyal-plane noncommutativity $\delta x_1 \delta x_2 \geq |\theta|/2$.
As already stressed above these relations do not obstruct the case of zero area,
because they allow the possibility that all points on the boundary sharply have
the same, say, $x_1$ coordinate. But they do exclude the case of zero distance:
for the distance between two points to be zero
one should demand that all coordinates of the two points
coincide, which is very clearly incompatible
with $\delta x_1 \delta x_2 \geq |\theta|/2$.

These observations suggest that, in spite of its relative simplicity,
the study of the Moyal plane may contribute some valuable indications
for the debate, within the quantum-gravity community, that indeed concerns
the possible emergence of features such as minimum area and minimum distance
at the Planck scale. In that debate a significant role is played by
semi-heuristic arguments that combine quantum mechanics and general relativity
and find rather robust (as robust as heuristic arguments can be)
evidence of the necessity to implement a Planckian minimum-distance principle.
Once the minimum-distance is (heuristically) established
 then it is not uncommon that, with somewhat looser use of logics,
 the authors
 make the assumption
that one would have to reach similar conclusions also for areas and volumes, as if in a quantum
spacetime the presence of a minimum-length bound would necessarily imply corresponding
bounds for areas and volumes. The Moyal plane, in light of the results we obtained,
is a quantum space that provides a counter-example for the assumptions
that guide this type of line of reasoning.

And it should be possible to investigate how common it is to find this specific scenario for
the spectra of distances and areas in noncommutative spaces. In fact,
many aspects of our analysis of the Moyal plane
are immediately adaptable to other, more complex, noncommutative spaces. In particular,
our formalization of the observable ``area of a triangle",
based on the novel
operator ${\hat A}({\vec {\hat X}}^{\, (1)},{\vec{\hat X}}^{\, (2)},{\vec {\hat X}}^{\, (3)})$,
should be applicable
without modification to a rather large class of noncommutative spaces.

\section*{Acknowledgments}
G.~A.-C. is supported by grant RFP2-08-02 from The Foundational Questions Institute (fqxi.org).\\
G.~G. is supported by ASI contract I/016/07/0 "COFIS".

\end{document}